\documentclass{aa}
  \usepackage{psfig}

\authorrunning{Brandner et al.}
\titlerunning{Detection of edge-on circumstellar disks in $\rho$ Oph}

\begin{document}

\thesaurus{08     
              (08.03.4;  
               08.06.2;  
               08.16.5;  
               10.15.2: rho Ophiuchi dark cloud)}  

\title{VLT-detection of two edge-on Circumstellar Disks 
in the $\rho$\,Oph dark cloud\thanks{Based on 
observations at the European Southern Observatory, Paranal
(ESO Prop ID 63.I-0691), the NAOJ SUBARU telescope at Mauna Kea, Hawai`i,
the ESA Infrared Space Observatory,
 and the NASA/ESA Hubble Space Telescope obtained at the Space 
Telescope Science Institute, which is operated by the Association of 
Universities for Research in Astronomy, Inc., under the NASA contract 
NAS5-26555.}}

\author{Wolfgang Brandner\inst{1}, 
Scott Sheppard\inst{1}
\and
Hans Zinnecker\inst{2}
\and
Laird Close\inst{3}
\and
Fumihide Iwamuro\inst{4}
\and
Alfred Krabbe\inst{5}\thanks{new address:
University of California at Berkeley, Department of 
Physics, 366 LeConte Hall \#7300, Berkeley, CA 94720-7300, USA}
\and
Toshinori Maihara\inst{4}
\and
Kentaro Motohara\inst{6}
\and
Deborah L.\ Padgett\inst{7}
\and
Alan Tokunaga\inst{1}
}

\institute{University of Hawaii, Institute for Astronomy, 2680 Woodlawn Dr., 
Honolulu, HI 96822, USA; brandner@ifa.hawaii.edu, sheppard@ifa.hawaii.edu,
tokunaga@ifa.hawaii.edu
\and
Astrophysikalisches Institut Potsdam, An der Sternwarte 16,
D-14482 Potsdam, Germany; hzinnecker@aip.de
\and
European Southern Observatory, Karl-Schwarzschild Stra{\ss}e 2, 
D-85748 Garching, Germany; lclose@eso.org
\and
Department of Physics, Kyoto University, Kitashirakawa, Kyoto 
606-8502, Japan; iwamuro@cr.scphys.kyoto-u.ac.jp, 
maihara@cr.scphys.kyoto-u.ac.jp
\and
Deutsches Zentrum f\"ur Luft- und Raumfahrt, Institut f\"ur Weltraumsensorik
und Planetenerkundung, Rutherfordstra{\ss}e 2, D-12489 Berlin, Germany;
krabbe@dlr.de
\and
Subaru Telescope, National Astronomical Observatory of Japan, 650 North
A`ohoku Place, Hilo, HI 96720, USA; motohara@subaru.naoj.org
\and
Jet Propulsion Laboratory, IPAC 100-22, California Institute of Technology, 
Pasadena, CA 91125, USA; dlp@ipac.caltech.edu
}

   \date{Received ; accepted}

   \maketitle

\begin{abstract} 
Observations of the $\rho$ Ophiuchi star forming region with VLT ANTU
and ISAAC under 0\farcs35 seeing conditions reveal two bipolar reflection
nebulosities intersected by central dust lanes. The sources 
(OphE-MM3 and CRBR 2422.8$-$3423) can be identified
as spatially resolved circumstellar  disks viewed close to
edge-on, similar to edge-on disk sources discovered previously
in the Taurus and Orion star forming regions. Millimeter continuum
fluxes yield disk masses of the order of
0.01\,M$_\odot$, i.e.\ about the mass deemed necessary for the minimum
solar nebula.  Follow-up spectroscopic observations with SUBARU and CISCO
show that both disk sources exhibit featureless continua in the K-band.
No accretion or outflow signatures were detected. The slightly
less edge-on orientation of the disk around CRBR 2422.8$-$3423 compared to HH 
30 leads to a dramatic difference
in the flux seen in the ISOCAM 4.5\,$\mu$m to
12\,$\mu$m bands. 
The observations confirm theoretical predictions on the effect of
disk geometry and inclination angle on the spectral energy distribution of
young stellar objects with circumstellar disks.

\end{abstract}


\section{Introduction}

One of the focal points of current astronomical research is the 
search for circumstellar disks and a study of their transformation into 
planetary systems. The aim is to understand how the solar system evolved,
and to determine how many other planetary systems might exist in the
Galaxy.

The huge brightness difference of 
typically 10$^6$ to 1 between the central 
star and light scattered from the disk surface (Sonnhalter et al.\ 1995;
Boss \& Yorke 1996) makes it challenging to
detect and resolve circumstellar disks in the optical and near-infrared.
If the disk, however, is seen close to edge-on, it acts as
a natural coronagraph and blocks out the light from the
central star. The typical signature for such an alignment is a
bipolar reflection nebula intersected by a central dust lane
(Whitney \& Hartmann 1992; Sonnhalter et al.\ 1995; Burrows et al.\ 1996).
Edge-on circumstellar disk sources have been discovered in
the Orion nebula (McCaughrean \& O'Dell 1996) and in the Taurus T\,association
(e.g., Burrows et al.\ 1996, Lucas \& Roche 1997, Padgett et al.\ 1999,
Koresko 1998, Stapelfeldt et al.\ 1998, Monin \& Bouvier 2000).

It has already become apparent that different environments affect
the evolution of circumstellar disks in various ways. In dense regions
with many hot, luminous early type stars like, e.g., the
Trapezium cluster in Orion or the starburst cluster NGC 3603, the 
harsh radiation environment leads to a rapid
photoevaporation of circumstellar disks (O'Dell et al.\ 1993;
McCullough et al.\ 1995; St\"orzer \& Hollenbach 1999; Brandner et al.\
2000). The low-density Taurus T\,association, on the other hand, is a more 
benign environment. Because of the high multiplicity among 
T\,Tauri stars
in Taurus (Ghez et al.\ 1993, Leinert et al.\ 1993, K\"ohler \& Leinert 1998),
however, the majority of circumstellar disks might be affected
by tidal truncation, like, e.g., HK Tau/c (Koresko 1998, Stapelfeldt et al.\ 
1998) or HV Tau C (Monin \& Bouvier 2000).

The $\rho$ Ophiuchi molecular cloud complex is the formation site
of a proto-open cluster with a gaseous mass
of around 550\,M$_\odot$ (Wilking \& Lada 1983) and at least
100 stellar members (Comer\'on et al.\ 1993; Barsony et al.\ 1997; Kenyon
et al.\ 1998). 
Its stellar density is intermediate between that of the Taurus
T\,association and the Trapezium cluster. The absence of hot, luminous
early-type stars ensures that circumstellar disks
are not subject to photoevaporation.
The $\rho$ Oph region appears to be a more typical representative of the
dominant star formation mode in the Galaxy than the low density
environment of the wide-spread Taurus T association with its high
percentage of binary and multiple systems.

In an effort to identify spatially resolved circumstellar disks
in various environments,
we carried out a VLT/ISAAC survey of southern starforming regions 
(Zinnecker et al.\ 1999). The survey also aimed at establishing a
sample of circumstellar disk sources which are suitable for
detailed follow-up studies with the VLT-Interferometer and the
Atacama Large Millimeter Array. 

\section{Observations and data reduction} 

\subsection{VLT/ISAAC imaging}

\subsubsection{Observations}

\begin{figure}[ht]
\centerline{\psfig{figure=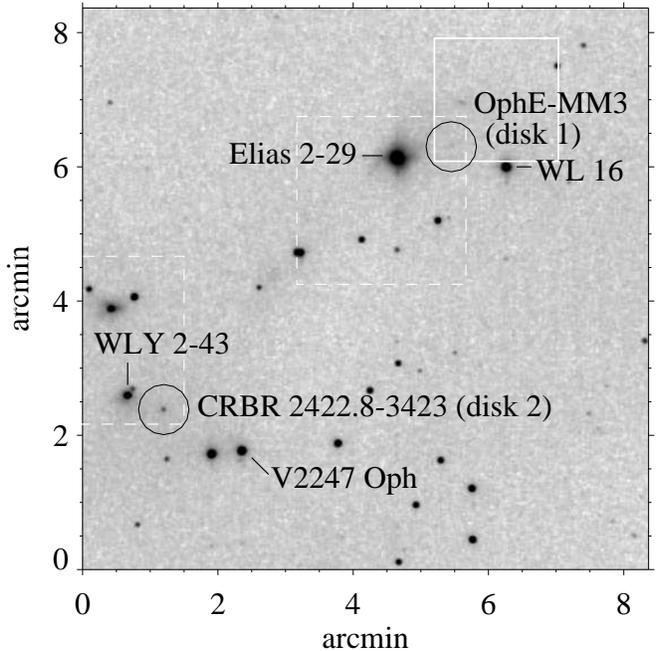,angle=0,width=9.2cm}}
\caption{K-band finding chart based on 2MASS atlas images.
The dashed lines indicate the area covered by the two VLT/ISAAC
pointings, and the solid lines mark the area imaged by SUBARU/CISCO.
The edge-on disk sources are located in the center of the black
circles.
\label{fig1}}
\end{figure}

Near-infrared J [1.11--1.39\,$\mu$m], H [1.50--1.80\,$\mu$m], and Ks 
[2.03--2.30\,$\mu$m] broad-band images of young stellar
sources in southern star forming regions were obtained on 28 April
1999 with the ESO VLT/UT1 (``Antu'') and the facility Infrared
Spectrometer And Array Camera (ISAAC, Moorwood et al.\ 1998). 
The observations were
carried out with the short-wavelength channel of ISAAC, which
is equipped with a 1024$\times$1024 Hawaii Rockwell HgCdTe array. The pixel
scale was 0\farcs147 pixel$^{-1}$ and the seeing 
was $\approx$0\farcs35 (FWHM) in Ks.  Two fields were centered
between Elias 29 and WL 19, and between WLY 43 and 44, respectively.
Total exposure times were 240s in H and 120s in Ks for the 
Elias 29/WL 19 field, and 640s in J, 320s in H, and 150s in Ks for the
WLY 43/44 field.  Figure \ref{fig1} illustrates the location
of the two ISAAC pointings.

\subsubsection{Data reduction and analysis}

\begin{figure}[htb]
\centerline{\psfig{figure=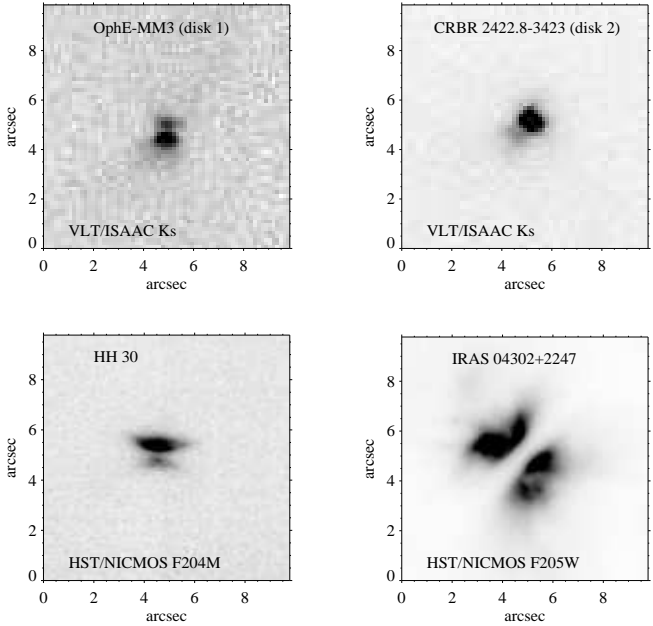,angle=0,width=8.6cm}}
\caption{Edge-on circumstellar disk sources in Ophiuchus and
Taurus. In comparison to isolated
disks in Taurus, the Ophiuchus disks and their reflection nebulosities
are more compact.
For the disk sources in the Ophiuchus region, North is up and East
is to the left.
\label{fig2}}
\end{figure}

Two bipolar reflection nebulosities intersected by dark lanes
were detected. The objects closely
resemble edge-on circumstellar disk sources in Taurus and Orion,
and in the following we refer to them as ``disk 1'' and ``disk 2''.
A close-up of the two new edge-on disk sources, and a
comparison to edge-on disks in Taurus is presented in Figure \ref{fig2}.
Absolute positions were
determined using 2MASS data products and relative offsets measured on
the ISAAC frames.
The photometric calibration is based on the observations
of infrared standard stars, and photometric measurements
from the 2MASS point source catalog.
Coordinates and near-infrared photometry of the disk sources are
summarized in Table \ref{tab1}.

\begin{figure}[htb]
\centerline{\psfig{figure=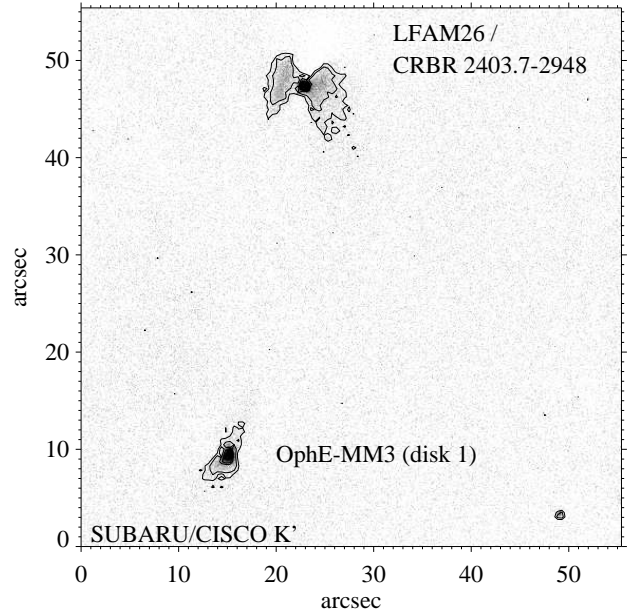,angle=0,width=8.5cm}}
\caption{Deep K' band imaging of disk 1 (OphE-MM3) 
reveals that the bipolar reflection
nebulosity stretches out for $\approx$5$''$ above and
below the plane of the disk. LFAM 26 (CRBR 2403.7$-$2948)
is associated with a bipolar reflection nebulosity intersected by
a central dust lane as well. Unlike disk 1 and 2, however,
a central point source is visible. North is up and East
is to the left.
\label{fig3}}
\end{figure}

\begin{table*}[htb]
\caption[]{Coordinates and near infrared photometry (in the 2MASS photometric
system) of the edge-on circumstellar disks.\label{tab1}} 
\begin{tabular}{lcccccc}
&$\alpha$(2000) &$\delta$(2000) &J & H & K & L$^{\mathrm{a}}$ \\
Target &[hms]   &[$\circ$ $'$ $''$] & [mag] &[mag] & [mag] &[mag]  \\ \hline
OphE-MM3 (disk 1) &16 27 05.91 &-24 37 08.2 &\dots    &18.4$\pm$0.2&15.4$\pm$0.2&\dots  \\
CRBR 2422.8$-$3423 (disk 2) &16 27 24.61 &-24 41 03.3 &$>$21.0    &18.0$\pm$0.2 & 13.3$\pm$0.1 & 9.7$\pm$0.1\\
\end{tabular}
\begin{list}{}{}
\item[$^{\mathrm{a}}$] Comer\'on et al.\ 1993
\end{list}
\end{table*}

\subsection{SUBARU/CISCO imaging and spectroscopy}

\subsubsection{Observations}

Long-slit K-band spectra were obtained on 18 June 2000 (``disk 2'') and
16 July 2000 (``disk 1'') with the SUBARU telescope
and the Cooled Infrared Spectrograph and Camera for OH suppression (CISCO,
Motohara et al.\ 1998). CISCO is equipped with a 1024$\times$1024 
Hawaii Rockwell HgCdTe array and has a pixel scale of 0\farcs111 pixel$^{-1}$. 
The slit-width was 1$''$ and the slit was oriented north-south.
Individual exposure
times were 60s and 100s, yielding a total integration time of 240s and 400s
for ``disk 2'' and ``disk 1'', respectively. 
Deep dithered K'-band [1.96--2.30\,$\mu$m] 
imaging data of disk 1 were obtained on 16 July 2000
(Figure \ref{fig3}). The seeing on the coadded 800s exposure is 0\farcs4.

\subsubsection{Data reduction and analysis}

The data reduction included sky subtraction, and a removal of the
geometrical distortion of the 2D spectra.
Telluric features were removed by dividing the spectrum of each disk by
the spectrum of the standard star.
As the observations were obtained under a relatively high airmass
of 1.6 to 2.0, the removal of the telluric features was not perfect.
The instrumental response was removed by applying an
appropriate library spectrum (Pickles 1998). The
resulting spectra are shown in Figure \ref{fig4}.
No variation in the spectra across the objects was seen, giving
further evidence that the objects are indeed reflection nebulosities.

\subsection{ISO observations}

The Ophiuchus region has been observed multiple times by ISO
(Kessler et al.\ 1996).
The regions including disks 1 and 2 were covered by pointed
ISOCAM (C\'esarsky et al.\ 1996) observations and by larger scale raster 
scans with ISOCAM and ISOPHOT (Lemke et al.\ 1996). HH 30 was also
observed by ISOCAM and ISOPHOT (PI: K.R.\ Stapelfeldt, see
Stapelfeldt \& Moneti 1999).
Basic science data were retrieved from the ISO archive and reprocessed
using the latest versions (as of June 2000) of the CIA and PIA software 
packages.

Disk 2 is detected in several ISOCAM observations towards
WLY 2-43 at wavelengths between 4.5\,$\mu$m and 11.3\,$\mu$m. 
Large area raster scans of the regions including disk 1 and disk 2
(Abergel et al.\ 1996) 
suffer from memory effects due to brighter sources in the vicinity of
the disk sources. It was therefore not possible to derive precise flux values
from the large area scans.
The ISOCAM fluxes of disk 2 and HH 30 are summarized in Table \ref{tab2}.

ISOPHOT maps at 60 and 100 $\mu$m detect and resolve Elias 2-29
and WLY 2-43. Disks 1 and 2, however, are located in the wings 
of the point spread function of the much brighter sources 
(FWHM for ISO is 20$''$ and 35$''$ at
wavelengths of 60\,$\mu$m and 100\,$\mu$m, respectively), and were not 
detected as individual sources.

\subsection{HST/NICMOS observations}

HST/NICMOS observations of HH 30 were retrieved from the HST archive. The 
data had been obtained as part of GTO 7228 (PI E.\ Young). IRAS 04302+2247
was studied by Padgett et al.\ (1999).
The NICMOS data were processed using the standard IRAF/STSDAS data
reduction pipeline, and the most current calibration files.

\section{Discussion}

\subsection{Disk 1 (OphE-MM3)}

The position of disk 1 agrees to within 1$''$ with the position of the
millimeter continuum source OphE-MM3 (Motte et al.\ 1998). This
millimeter source was detected as part of a detailed 1.3mm continuum 
survey of the Ophiuchus star forming region. Because of the lack
of any known counterpart shortward of 1.3\,mm, Motte et al.\ (1998) classify
OphE-MM3 as a compact starless core. 
Motte et al.\ (1998) note that the millimeter continuum emission of OphE-MM3 
is more centrally peaked and does not exhibit any inner flattening as other
starless cores in their sample.

The near-infrared counterpart to OphE-MM3 is easily detected on
the VLT/ISAAC observations in H and Ks. The high-spatial resolution VLT data
reveal that the relative faintness of OphE-MM3 is due to the orientation
of its circumstellar disk. The disk is seen close to edge-on, and appears
as a central absorption band between the bipolar reflection nebulosity
(Figure \ref{fig2}).
The Ks-band peak surface brightness ratio between the southern and the 
northern half of the reflection nebulosity is 2.1 to 1.
OphE-MM3 can also be identified as a 3$\sigma$ detection on the 
2MASS K-band atlas data. Because of its proximity to the K=6\fm8 source 
Elias 2-29 (10\farcs49 south and 48\farcs08 east of OphE-MM3), and the 
increased local background due to diffuse near-infrared 
emission, however, it must have been overlooked in previous infrared surveys of 
the $\rho$\,Oph region. 
The deeper SUBARU/CISCO K'-band images of OphE-MM3 reveal that the
bipolar reflection nebulosity extends for $\approx$5$''$ above and below the
plane of the disk (Figure \ref{fig3}). The K-band spectrum of OphE-MM3
does not show any strong emission lines (Figure \ref{fig4}).

For young stellar objects, the mass in the disk and the infalling envelope
can be derived from 1.3\,mm continuum observations (Beckwith et al.\ 1990).
Motte et al.\ (1998) measure a peak flux beam$^{-1}$ of 60$\pm$10\,mJy
for OphE-MM3.  Assuming the standard set of
parameters, i.e., T=30\,K, $\kappa_{\rm 1.3mm}$=0.01\,cm$^2$g$^{-1}$,
a gas to dust ratio of $\approx$100, and a distance of 150\,pc yields a mass
of 0.015\,M$_\odot$, i.e., about the same mass typically assumed for
the ``minimum solar nebula'' (e.g., Yorke et al.\ 1993).

\subsection{Disk 2 (CRBR 2422.8$-$3423)}

This source has been detected in previous near-infrared 
observations (CRBR 2422.8$-$3423 -- Comer\'on et al.\ 1993; [SKS95] 
162422.8-243422 -- Strom et al.\ 1995; BKLT J162724$-$244103 -- Barsony et al.\ 
1997) and is included in the 2MASS point source catalog. 
CRBR 2422.8$-$3423 is located 12\farcs72 south and 
31\farcs75 west of WLY 2-43.
Similar to OphE-MM3, the VLT/ISAAC observations reveal
CRBR 2422.8$-$3423 as a bipolar nebulosity intersected by a central dust
lane. The source is detected on the ISAAC H- and Ks-band frames, but not
in the J-band. The magnitudes derived from the VLT/ISAAC and 2MASS data 
are H=18\fm0 and Ks=13\fm3. The Ks-band magnitude is in good agreement
to the measurements by Comer\'on et al.\ (1993) and Barsony et al.\
(1997). In Ks, the peak surface brightness ratio between the north-western 
and the south-eastern halves of the reflection nebulosity is 11 to 1. The 
larger brightness ratio indicates that the disk plane of CRBR 2422.8$-$3423 is 
farther from an edge-on inclination than the disk of OphE-MM3.

\begin{figure}[htb]
\centerline{\psfig{figure=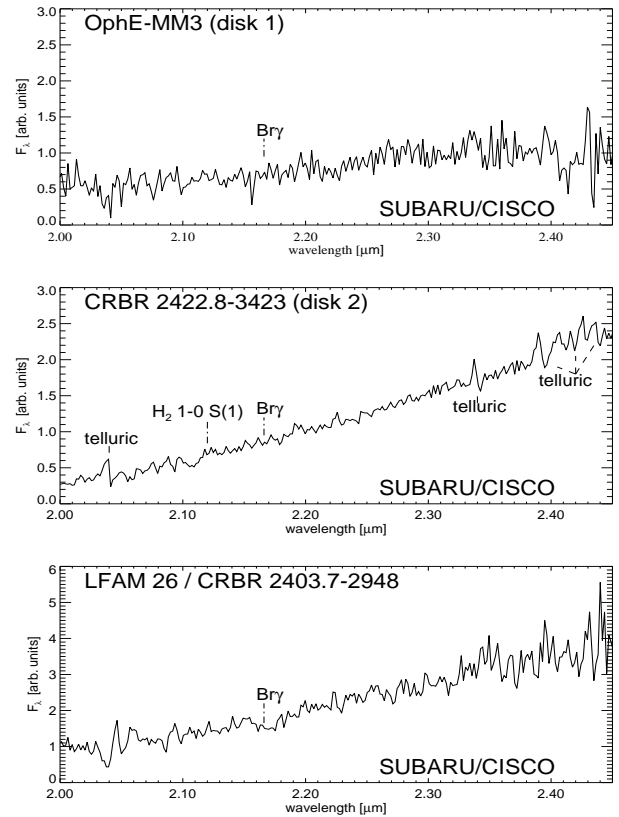,angle=0,width=8.6cm,height=10.2cm}}
\caption{K-band spectra of disk 1, disk 2, and LFAM26 obtained with
SUBARU/CISCO.
All three sources exhibit a featureless continuum and show no
signs of underlying stellar photospheres, or accretion and outflow signatures
(e.g., H$_2$ or
Br$\gamma$ emission).
\label{fig4}}
\end{figure}

CRBR 2422.8$-$3423 has very red near-infrared colours (Table \ref{tab1}). 
The steeply rising continuum towards longer wavelengths is also apparent in the
K-band spectrum (Figure \ref{fig4}). No signs of an underlying stellar 
photosphere or accretion and outflow signatures, like H$_2$ or Br$\gamma$
emission lines, are present. There is also no evidence 
of a CO band at 2.3\,$\mu$m.
The featureless K-band continuum suggests that the observed K-band radiation
is largely due to reprocessed photons from hot dust near the star.

\begin{table*}[htb]
\caption[]{ISOCAM fluxes of edge-on disk sources.\label{tab2}}
\begin{tabular}{lcccccccc}
 &4.5\,$\mu$m  &6.0\,$\mu$m  &6.8\,$\mu$m  & 7.7\,$\mu$m  & 9.6\,$\mu$m  & 11.3\,$\mu$m  & 12\,$\mu$m & 14.3\,$\mu$m\\
Target &[mJy]   &[mJy] & [mJy] &[mJy] & [mJy] &[mJy] &[mJy] &[mJy] \\ \hline
CRBR 2422.8$-$3423 &260$\pm$30 &400$\pm$30 &400$\pm$40 &410$\pm$10 &150$\pm$20 &330$\pm$30 &\dots &\dots \\
HH 30 & 3.0$\pm$0.8 &\dots &2.1$\pm$0.8 &\dots &2.3$\pm$0.8 &\dots &2.7$\pm$0.8 &3.9$\pm$0.8 \\
\end{tabular}
\end{table*}

Motte et al.\ (1998) identify CRBR 2422.8$-$3423 as an unresolved 1.3\,mm
continuum source with a peak flux beam$^{-1}$ of 40$\pm$10\,mJy.
This corresponds to a mass of 0.01\,M$_\odot$ for the
disk and the infalling envelope, again in good agreement with the
mass of the ``minimum solar nebula''.

\subsection{LFAM 26 (CRBR 2403.7$-$2948)}

The deep SUBARU/CISCO K'-band imaging reveals another source with a bipolar
reflection nebulosity $\approx$40$''$ to the north of disk 1 
(Figure \ref{fig3}). This object has
previously been detected at infrared and radio wavelengths
(LFAM 26 -- Leous et al.\ 1991;
[GY92] 197 -- Greene \& Young 1992; CRBR 2403.7$-$2948 -- Comer\'on et al.\ 1993;
[SKS95] 162403.8-242948 -- Strom et al.\ 1995).
Motte et al.\ (1998) measure a peak flux beam$^{-1}$ of 75$\pm$10\,mJ 
at 1.3\,mm.
The disk/envelope around LFAM 26 thus appears to be the most massive
among the three Ophiuchus sources discussed in this letter.
The wide dust lane and the fact that a central point source is visible
distinguishes this source from the two edge-on disks.

\subsection{Comparison to HH 30 and IRAS04302+2247}

The two best studied examples for edge-on circumstellar disks are HH 30
(Burrows et al.\ 1996, Stapelfeldt et al.\ 1998) and IRAS 04302+2247
(Lucas \& Roche 1998, Padgett et al.\ 1999).
With a 1.3\,mm continuum flux of 35\,mJy (Reipurth et al.\ 1983),
the amount of circumstellar dust in HH 30 is comparable to
CRBR 2422.8$-$3423, and somewhat less than in OphE-MM3. The disk and envelope
of IRAS 04302$+$2247 (F$_\nu$ = 150 mJy at 1.1\,mm, Hogerheijde et al.\ 1997)
is 3 to 4 times more massive.

The peak surface brightness ratio of the two parts of the bipolar reflection
nebulosity of HH 30
in the HST/NICMOS F204M filter is 3.7 to 1, i.e., intermediate between
OphE-MM3 and CRBR 2422.8$-$3423. The inclination angle
of the disk of HH 30 with respect to the line of sight
as derived by Burrows et al.\ (1996) is $\le$10$^\circ$.

Figure \ref{fig2} illustrates the morphology of four 
edge-on circumstellar disk sources.
IRAS 04302$+$2247 is more extended and appears to be in an earlier
evolutionary stage than the `bare-disk' system HH 30.  Similar to HH 30, 
the disk of OphE-MM3 exhibits a hint of flaring, whereas the central dust
lane of CRBR 2422.8$-$3423 is flat.  The extended reflection nebulosities
outline the interface between an outflow cavity and the infalling
circumstellar envelope (see, e.g., Wilkin \& Stahler 1998).
Like IRAS 04302$+$2247, OphE-MM3 and CRBR 2422.8$-$3423 are more deeply embedded 
and thus appear to be younger than
HH 30, which has already lost most of its envelope material.

\begin{figure}[htb]
\centerline{\psfig{figure=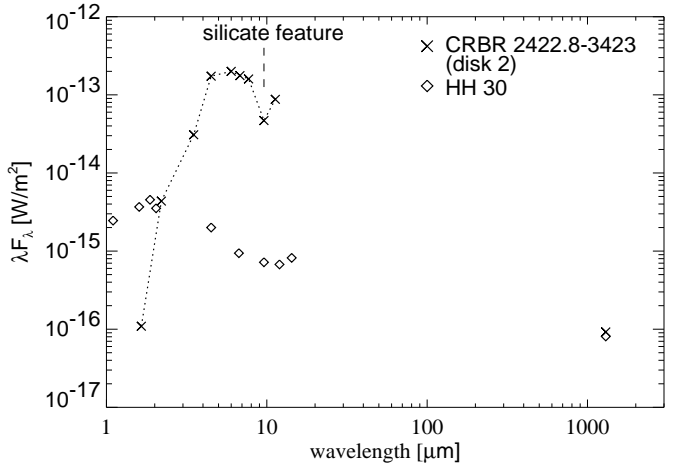,angle=0,width=8.5cm}}
\caption{Spectral energy distribution of CRBR 2422.8$-$3423 (disk 2)
and HH 30. Both sources exhibit about the same flux at 2.2\,$\mu$m and
1.3\,mm. The slightly larger inclination of CRBR 2422.8$-$3423's disk
allows the warm, inner disk to become detecable at NIR to MIR wavelengths.
The dip at 9.6\,$\mu$m can be explained by absorption due to the
silicate feature. The spectral energy distribution of HH 30, whose disk is
seen closer to edge-on than in the case of CRBR 2422.8$-$3423,
is dominated by scattered light out to wavelengths of 10\,$\mu$m.
\label{fig5}}
\end{figure}

Figure \ref{fig5} displays the spectral energy distribution (SED)
of CRBR 2422.8$-$3423
and HH 30 for wavelengths between 1\,$\mu$m and 1.3\,mm.  While
both sources have very similar K band brightness and
exhibit about the same 1.3\,mm continuum flux, the overall shape of the
SED is vastly different. The SED of CRBR 2422.8$-$3423 is rising steeply in the 
NIR,
peaks at around 5\,$\mu$m, and is slowly declining towards longer
wavelengths. A sharp dip in the SED is apparent at 9.6\,$\mu$m. 
The data confirm the classification of CRBR 2422.8$-$3423 as a Class 
I source as suggested by Motte et al.\ (1998). 
The SED of HH 30 peaks at around 2\,$\mu$m and shows a
shallow and very broad dip around 10\,$\mu$m.

As a preparation for the ISO and SIRTF missions,
Boss \& Yorke (1996)
carried out extensive radiative hydrodynamical simulations 
to study the effect of disk geometry and physical parameters
like disk mass, stellar mass, dust opacity, or accretion rate
on the SED of circumstellar disks.

The inclination of a disk to the line of sight has a strong impact
on the overall SED. Close to edge-on,
a disk source is almost entirely seen in scattered light out to
wavelengths of 20\,$\mu$m. Edge-on orientation also causes the
ice band at 3\,$\mu$m and the silicate feature at 10\,$\mu$m to be
seen in emisison. Millimeter fluxes are hardly affected by the disk's
inclination.
For a less edge-on inclination the circumstellar envelope and the
outer parts of the disk become optically thin in the mid-infrared, and continuum
emission from the warm, inner disk is detected. As a consequence, the
silicate feature can be seen in absorption. The simulations by Boss \& Yorke
(1996) also indicate that the silicate absorption feature becomes more
pronounced with increasing disk mass, as a higher disk mass leads to a higher
central plane temperature of the inner disk.

The simulations nicely explain the SED of
the edge-on disk sources CRBR 2422.8$-$3423 and HH 30. HH 30 is seen
so close to edge-on that only scattered light is detected out to
a wavelength of 10\,$\mu$m. Both the faintness of HH 30 and the limited
spectral resolution of the ISOCAM bands might contribute to a non-detection
of a silicate emission feature.
The disk of CRBR 2422.8$-$3423 is oriented less edge-on, and the warm, inner 
disk can be detected in the mid-infrared.
The sharp dip at 9.6\,$\mu$m in the SED of CRBR 2422.8$-$3423 is due
to silicates seen against the continuum emission from the warm, inner disk.

The observations thus confirm theoretical predictions on the
effect of disk geometry and disk inclination on the spectral appearance
of a young stellar object.

\section{Summary}

We have presented the first identification of edge-on circumstellar disks
outside of the Orion and Taurus star forming regions. Compared to
the majority of edge-on circumstellar disk sources in Taurus, the two 
edge-on disks in the Ophiuchus region appear to be at an earlier
evolutionary stage, as they are still more deeply embedded in their
circumstellar envelopes. The high-spatial resolution infrared data
confirm that the difference in the spectral energy distribution
of CRBR 2422.8$-$3423 and HH 30 is due to a slight difference in the
inclination of the disks to the line of sight, as had been predicted
previously by theoretical models.

\begin{acknowledgements}
We would like to thank the SUBARU telescope and VLT staff members
for their excellent work, and the referee Dr.\ K.R.\ Stapelfeldt
for his critical and helpful comments.
This publication makes use of data products from 2MASS, which is a
joint  project of UMass and IPAC/Caltech, funded by NASA and NSF.
WB acknowledges support by the National Science Foundation and by NASA. 
\end{acknowledgements}


\begin{thebibliography}{}

\bibitem{} Abergel, A., Bernard, J.P., Boulanger, F., C\'esarsky, C., D\'esert, 
F.X.\ et al.\ 1996, A\&A 315, L329

\bibitem{} Barsony, M., Kenyon, S.J., Lada, E.A., Teuben, P.J.\ 1997, ApJS 112,
109

\bibitem{} Beckwith, S.V.W., Sargent, A.I., Chini, R.S., Guesten, R.\ 1990, AJ,
99, 924

\bibitem{} Boss, A.P., Yorke, H.W.\ 1996, ApJ, 469, 366

\bibitem{} Brandner, W., Grebel, E.K., Chu, Y.-H., Dottori, H., Brandl, B.,
Richling, S., Yorke, H.W., Points, S.D., Zinnecker, H.\ 2000, AJ 119, 292

\bibitem{} Burrows, C.J., Stapelfeldt, K.R., Watson, A.M., Krist, J.E., Ballester, G.E.\ et al.\ 1996, ApJ 473, 437

\bibitem{} C\'esarsky, C.J., Abergel, A., Agn\`ese, P.\ et al.\ 1996, A\&A 315, 
 L32

\bibitem{} Comer\'on, F., Rieke, G.H., Burrows, A., Rieke, M.J.\ 1993, ApJ 416,
185

\bibitem{} Ghez, A.~M., Neugebauer, G., Matthews, K. 1993, AJ 106, 2005

\bibitem{} Greene, T.P., Young, E.T.\ 1992, ApJ 395, 516

\bibitem{} Kenyon, S.J., Lada, E.A., Barsony, M.\ 1998, AJ 115, 252

\bibitem{} Kessler, M.F., Steinz, J.A., Anderegg, M.E., Clavel, J., 
Drechsel, G.\ et al.\ 1996, A\&A 315, L27

\bibitem{} K\"ohler, R., Leinert, Ch. 1998, A\&A 331, 977

\bibitem{} Koresko, C.D.\ 1998, ApJ 507, L145

\bibitem{} Leinert, Ch., Zinnecker, H., Weitzel, N.,
        Christou, J., Ridgway, S.~T., Jameson, R., Haas, M.,
        Lenzen, R.  1993, A\&A 278, 129

\bibitem{} Lemke, D., Klaas, U., Abolins, J.\ et al.\ 1996, A\&A 315, L64

\bibitem{} Leous, J.A., Feigelson, E.D., Andr\'e, P., Montmerle, T.\ 1991, ApJ
379, 683

\bibitem{} Lucas, P.W., Roche, P.F.\ 1997, MNRAS 286, 895

\bibitem{} McCaughrean, M.J., O'Dell, C.R.\ 1996, AJ, 111, 1977

\bibitem{} McCullough, P.R., Fugate, R.Q., Christou, J.C., Ellerbroek, B.L.,
Higgins, C.H., Spinhirne, J.M., Cleis, R.A., Moroney, J.F.\ 1995, ApJ 438, 394

\bibitem{} Monin, J.-L., Bouvier, J.\ 2000, A\&A 356, L75

\bibitem{} Moorwood, A., Cuby, J.-G., Biereichel, P., Brynnel, J.,
Devillard, N., et al.\ 1998, The Messenger, 94, 7

\bibitem{} Motohara, K., Maihara, T., Iwamuro, F., Oya, S., Imanishi, M.,
Terada, H., Goto, M., Iwai, J., Tanabe, H., Tsukamoto, H., Sekiguchi, K.\
1998, Proc.\ SPIE, 3354, 659

\bibitem{} Motte, F., Andr\'e, P., Neri, R.\ 1998, A\&A 336, 150

\bibitem{} O'Dell, C.R.O., Wen, Z., Hu, X.\ 1993, ApJ, 410, 686

\bibitem{} Padgett, D.L., Brandner, W., Stapelfeldt, K.R., Strom, S.E.,
Terebey, S., Koerner, D.\ 1999, AJ 117, 225

\bibitem{} Pickles, A.J.\ 1998, PASP 110, 863

\bibitem{} Reipurth, B., Chini, R., Kr\"ugel, E., Kreysa, E., Sievers, A.\ 1993, A\&A  273, 221

\bibitem{} Sonnhalter, C., Preibisch, T., Yorke, H.W.\ 1995, A\&A 299, 545

\bibitem{} Stapelfeldt, K.R., Krist, J.E., Menard, F., Bouvier, J., Padgett,
D.L., Burrows, C.J.\ 1998, ApJ 502, L65

\bibitem{} Stapelfeldt, K.R., Moneti, A.\ 1999, in {\it The Universe as Seen 
by ISO}, eds.\ P.\ Cox \& M.F.\ Kessler, ESA-SP 427, p.\ 521

\bibitem{} St\"orzer, H., Hollenbach, D., 1999, ApJ, 515, 669

\bibitem{} Strom, K.M., Kepner, J., Strom, S.E.\ 1995, ApJ 438, 813

\bibitem{} Whitney, B.A, Hartmann, L.\ 1992, ApJ 395, 529

\bibitem{} Wilkin, F.P., Stahler, S.W.\ 1998, ApJ 502, 661

\bibitem{} Wilking, B.A, Lada, C.J.\ 1983, ApJ 274, 698

\bibitem{} Yorke, H.W., Bodenheimer, P., Laughlin, G.\ 1993, A\&A 411, 274

\bibitem{} Zinnecker, H., Krabbe, A., McCaughrean, M.J., Stanke, T., 
Stecklum, B., Brandner, W., Padgett, D.L., Stapelfeldt, K.R., Yorke, H.W.\
1999, A\&A 352, L73

\end{thebibliography}
\end{document}